\documentclass[prd,twocolumn,floatfix,preprintnumbers]{revtex4}

\usepackage{graphicx}
\usepackage{graphicx,epsfig}
\usepackage{bm}
\usepackage{latexsym,amssymb,amsmath,float}

\newcommand {\ga} {\ {\raise-.5ex\hbox{$\buildrel>\over\sim$}}\ }
\newcommand {\la} {\ {\raise-.5ex\hbox{$\buildrel<\over\sim$}}\ }

\def\be{\begin{equation}}
\def\ee{\end{equation}}
\def\ba{\begin{eqnarray}}
\def\ea{\end{eqnarray}}

\renewcommand{\(}{\left(}
\renewcommand{\)}{\right)}
\renewcommand{\[}{\left[}
\renewcommand{\]}{\right]}

\begin{document}

\title{Coincidence Problem in Cyclic Phantom Models of the Universe}
\author{Hui-Yiing Chang and Robert J. Scherrer}
\affiliation{Department of Physics and Astronomy, Vanderbilt University,
Nashville, TN  ~~37235}

\begin{abstract}
We examine cyclic phantom models for the universe, in which
the universe is dominated sequentially by radiation, matter, and a phantom dark energy field,
followed by a standard inflationary phase.  Since this
cycle repeats endlessly, the Universe
spends a substantial portion of its lifetime in a state for which the matter and dark energy
densities have comparable magnitudes, thus
ameliorating the coincidence problem.  We calculate the fraction of
time that the universe spends in such a coincidental state and find that it is nearly the
same as in the case of a phantom model with a future big rip.  In the limit
where the dark energy equation of state parameter, $w$, is close to $-1$, we show that the fraction of time, $f$, for which
the ratio of the dark energy density to the matter density lies between $r_1$ and $r_2$, is
$f = -(1+w) \ln [(\sqrt{r_2} + \sqrt{1+r_2})/(\sqrt{r_1} + \sqrt{1+r_1})].$
\end{abstract}

\maketitle

\section{Introduction}

Cosmological data \cite{union08,perivol,hicken,Komatsu}
indicate that approximately
70\% of the energy density in the
universe is in the form of an exotic, negative-pressure component,
called dark energy, with roughly 30\% in the form of nonrelativistic matter (including both baryons
and dark matter).
The dark energy component can be parametrized by its equation of state parameter, $w$,
defined as the ratio of the dark energy pressure to its density:
\be
\label{w}
w=p_{\rm DE}/\rho_{\rm DE},
\ee
where $w=-1$ corresponds to a cosmological constant.
For constant $w$, the energy density of the dark energy, $\rho_{DE}$, scales as
\be
\label{rhode}
\rho_{DE}=\rho_{DE0}\(\frac{R}{R_0}\)^{-3(1+w)},
\ee
where $R$ is the scale factor, and $\rho_{DE0}$ and
$R_0$ are the density and scale factor, respectively, at the present.
(We will use zero subscripts throughout to refer to present-day values). 
Observations constrain $w$ to be very close to $-1$.
For example, if $w$ is assumed to be constant, then $-1.1 \la w \la -0.9$  \cite{Wood-Vasey,Davis}.
Thus, the dark energy density varies relatively slowly with scale factor.

The matter density, in contrast, scales as
\be
\label{rhom}
\rho_M = \rho_{M0}\(\frac{R}{R_0}\)^{-3}.
\ee
This leads to the well-known coincidence problem:  while the matter and dark energy densities
today are nearly within a factor of two of each other, at early times $\rho_M \gg \rho_{DE}$,
and in the far future we expect $\rho_{DE} \gg \rho_M$.  It would appear, then, that we live in a 
very special time:  this is the coincidence problem.

While it is possible that this coincidence has no
deeper explanation, numerous solutions have been proposed to explain it.  In
the $k$-essence model of Armendariz-Picon et al. \cite{Armendariz-Picon},
the dark energy density tracks the
radiation density during the radiation-dominated
epoch but approaches a constant value during
the matter-dominated epoch.  Another proposed
solution is a universe which experiences an alternation
of matter domination and dark energy domination, either through
a scalar field with oscillatory behavior \cite{Dodelson,Feng}, or as a result of
a variety of scalar fields with a wide
range of energy densities \cite{Griest}.  Another possible solution for
the coincidence problem is a coupling of the matter and quintessence
fields so that energy is transferred between them \cite{Huey,Chimento}.
Garriga and Vilenkin \cite{Garriga} proposed an anthropic solution
to the coincidence problem. Scherrer \cite{Scherrer}
suggested that the coincidence
problem could be resolved in the context of phantom dark energy models.
In such models, the universe terminates in a singularity at a finite time \cite{Caldwell2,Caldwell3},
so that the fraction of time for which the dark energy and matter densites are relatively close
can be a significant fraction of the universe's (finite) lifetime.
Other models in which the coincidence problem is resolved by the universe having
a finite lifetime were examined by Barreira and Avelino \cite{BA}.
Lineweaver and Egan \cite{Lineweaver1,Lineweaver2} have proposed
that the coincidence is related to the formation rate for habitable planets.  

Here we examine another plausible solution to the coincidence problem, in the context of
cyclic phantom models, of the type proposed by
Ilie et al. \cite{Ilie}.  In these models,
the universe goes through repeated
cycles of matter/radiation domination followed by a dark energy/inflationary phase.
Ilie et al. indicated
that their model cannot address the coincidence problem, but we show
here that it provides an elegant resolution of this problem.
Within
each cycle, there is a significant period in which the dark energy
and matter densities are comparable.  Since these cycles repeat endlessly,
it is not surprising that we find ourselves in an epoch in which
the dark energy and matter densities are of the same order of magnitude.
Similar models have been proposed by Creminelli et al. \cite{Crem} and
by Xiong et al. \cite{Xiong}.

We make this argument quantitative in the next section.  Rather than
confining ourselves to the specific model of Ref. \cite{Ilie}, we use a toy
model which captures the essential features of a generic cyclic phantom model.
We also derive a
useful approximation to the coincidence fraction in the limit where $w$
is close to $-1$ (as observations require).  Our results are discussed
in Sec. III.

\section{The Coincidence Fraction in the cyclic phantom Model}

In the cyclic phantom model proposed by Ilie et al. \cite{Ilie}, the universe
contains radiation, a scalar field, and a hidden matter sector.
Inflationary expansion is followed by a reheating phase, during which radiation
becomes the dominant component.  Eventually the
scalar field and hidden matter densities both track the radiation density
but are subdominant.  At late times, the hidden matter and scalar field begin to behave as
a phantom field with $w < -1$, and the universe undergoes superaccelerated expansion.  This
phase then transitions to de Sitter inflation, and the cycle repeats itself.

Much of the complexity of the model discussed in Ref. \cite{Ilie} stems from the need to have a plausible
mechanism for the universe to transition from one phase of the expansion to the next.  Since we
are primarily interested in the behavior of the scale factor
as a function of time, we will consider
a toy model that approximates the general behavior of a cyclic phantom model. (This is
also necessitated by the fact that the model introduced in Ref. \cite{Ilie}
does not contain a matter component).  The use of such
a toy model has the additional advantage of being applicable to more general cyclic phantom models
than the specific model in Ref. \cite{Ilie}.
(As we have already noted, a number of similar models have been proposed \cite{Crem,Xiong}).

In our toy model, the universe undergoes an initial ``standard" expansion, consisting of a
radiation-dominated era, followed by a matter-dominated era.  An additional dark
energy component is present,
which tracks the matter or radiation density, but which is subdominant (so
that $\rho_{DE} \ll \rho_M$ in the matter-dominated era).
When the dark energy
density reaches some lower energy scale $m$ (so that $\rho_{DE} \sim m^4$), the
dark energy assumes a phantom behavior, with equation
of state parameter $w < -1$, and the universe undergoes superaccelerated expansion.  This phantom
phase terminates when the dark energy density (which is increasing with the expansion) reaches
some upper energy scale $M$, so that $\rho_{DE} \sim M^4$.  The universe then enters a de Sitter phase, which
ends with reheating and a return to the radiation-dominated era.

The solution to the coincidence problem in this model arises because the universe naturally spends
a significant fraction of the time in a state in which the densities of the dark energy
and the matter are of the same order of magnitude.  Conceptually, then, this solution resembles that
of Dodelson et al. \cite{Dodelson}, in which the ratio of dark energy density
to the density of the matter/radiation component oscillates with time.  Mathematically, however,
it more closely resembles the discussion in Ref. \cite{Scherrer} for models with a single
phantom phase terminating in a big rip, and it is this latter approach which we will
follow in analyzing the cyclic phantom model.

Our goal is to derive the fraction of the time that the universe spends
in a coincidental state, defined to be a state for which the ratio of
the density of dark energy to the density of matter lies within some fixed range
close to 1.  More
specifically, let $\rho_{DE}$ be the dark energy density, and $\rho_{M}$ be the nonrelativistic
matter density, and define the ratio $r$ as in Ref. \cite{Scherrer}:
\begin{equation}
r=\frac{\rho_{DE}}{\rho_{M}}.
\end{equation}
We will then define a coincidental state to be one for which $r$ lies in the range
\begin{equation}
r_1 < r < r_2,
\end{equation}
where the values for $r_1$ and $r_2$ that define
a ``coincidence" are, of course, somewhat arbitrary.

We assume a flat Friedman-Robertson-Walker model, so that the
evolution of the scale factor is given by
\be
\(\frac{\dot {R}}{R}\)^{2}=\frac{8}{3} \pi G \rho.
\ee
At late times,
the expansion of the universe is dominated by matter and dark energy.
To simplify matters, we assume throughout that $w$ is constant.  Then
we can use Eqs. (\ref{rhode}) and (\ref{rhom}) to give
\be
\(\frac{\dot {R}}{R}\)^{2}=\frac{8}{3} \pi G \[\rho_{M0}\(\frac{R}{R_0}\)^{-3}+\rho_{DE0}\(\frac{R}{R_0}\)^{-3(1+w)}\].
\ee
The time the universe takes in expanding from scale factor $R_1$ to $R_2$ is
\ba
t_{12}&=&\int_{R_1}^{R_2} R^{-1} \Bigg\{\frac{8}{3} \pi G \Bigg[\rho_{M0}\(\frac{R}{R_0}\)^{-3}
\nonumber\\
\label{t12}
&&+\rho_{DE0}\(\frac{R}{R_0}\)^{-3(1+w)}\Bigg]\Bigg\}^{-\frac{1}{2}}\, dR,
\ea
and the time the universe takes to complete one cycle is
\begin{eqnarray}
t_{cycle}&=&\int_{R = 0}^{R=R_{max}} 
R^{-1} \Bigg\{\frac{8}{3} \pi G \Bigg[\rho_{M0}\(\frac{R}{R_0}\)^{-3}
\nonumber\\
\label{cycle}
&&+\rho_{DE0}\(\frac{R}{R_0}\)^{-3(1+w)}\Bigg]\Bigg\}^{-\frac{1}{2}}\, dR,
\end{eqnarray}
where $R_{max}$ is the scale factor at which the dark energy density reaches its
maximum value of $M^4$ and the de Sitter phase begins.
Note that the integrand in equation (\ref{cycle}) is valid only after the dark
energy begins to behave as a phantom, but the error involved in extrapolating it back
to $R=0$ is negligible.

The fraction of time in each cycle that the universe spends in expanding from $R_1$ to $R_2$ is
$f=t_{12}/t_{cycle}$.
As in ref. \cite{Scherrer}, we can rewrite $t_{12}$ and $t_{cycle}$ in terms of $r$.
Taking
$r_1$ to be the value of $r$ at the beginning of the period of coincidence and $r_2$ as that at
the end, the fraction of time in each cycle that the universe spends in a coincidental state is
\be
\label{fraction}
f=\frac{\int_{r_1}^{r_2} {r^{-\frac{2w+1}{2w}}}/{\sqrt{1+r}}\,
dr}{\int_{0}^{r= M^4/\rho_{Mmax}} {r^{-\frac{2w+1}{2w}}}/{\sqrt{1+r}}\,
dr},
\ee
where $\rho_{Mmax}$ is the value of the matter density at $R_{max}$.
Since the cycles are identical and repeat indefinitely, $f$ is also the fraction of the entire universe's lifetime that is spent in a coincidental state.

This coincidence fraction is at least as large as in the case of a future big rip singularity
\cite{Scherrer}, and in principle it can be even larger, since the upper limit in the denominator
of equation (\ref{fraction}) is finite in the case considered here.  This upper limit is enormous, but the integral
converges very slowly for $w$ near $-1$, so it is useful to see how small
$M^4$ needs to be in order for the result to diverge significantly
from the case investigated in Ref. \cite{Scherrer}.
We have
\be
\rho_{DE0} \bigg(\frac{R_{max}}{R_0}\bigg)^{-3(1+w)}=M^4.
\ee
Therefore, the matter density at $R_{max}$ can be expressed as
\be
\rho_{Mmax}=\rho_{M0} \bigg(\frac{R_{max}}{R_0}\bigg)^{-3}=\rho_{M0}
\bigg(\frac{M^4}{\rho_{DE0}}\bigg)^{\frac{1}{1+w}}.
\ee
This allows us to express the upper limit of integration in the denominator
of Eq. (\ref{fraction}) as
\be
\label{rlimit}
r=\frac{M^4}{\rho_{Mmax}}=\bigg(\frac{\rho_{DE0}}{\rho_{M0}}\bigg)
\bigg(\frac{M^4}{\rho_{DE0}}\bigg)^{\frac{w}{1+w}}.
\ee
At the present, we have $\rho_{M0} \sim \rho_{DE0}$.  Using this in
Eq. (\ref{rlimit}), we can rewrite Eq. (\ref{fraction}) as
\be
\label{fraction2}
f=\frac{\int_{r_1}^{r_2} {r^{-\frac{2w+1}{2w}}}/{\sqrt{1+r}}\,dr}
{\int_{0}^{r= (M/E_{DE0})^{4w/(1+w)}} {r^{-\frac{2w+1}{2w}}}/{\sqrt{1+r}}\,
dr},
\ee
where the present-day energy scale of the dark energy is $E_{DE0} \sim
10^{-3}$ eV, and $\rho_{DE0} = E_{DE0}^4$.
The denominator in Eq. (\ref{fraction2}) can be expressed as
\begin{widetext}
\begin{eqnarray}
\int_{0}^{r= (M/E_{DE0})^{4w/(1+w)}} \frac{r^{-\frac{2w+1}{2w}}}{\sqrt{1+r}} dr
&= &\int_{0}^\infty \frac{r^{-\frac{2w+1}{2w}}}{\sqrt{1+r}}dr
 - \int_{r= (M/E_{DE0})^{4w/(1+w)}}^\infty \frac{r^{-\frac{2w+1}{2w}}}{\sqrt{1+r}} dr,\nonumber\\
\label{denom}
&\approx& \frac{\Gamma(-1/2w)\Gamma(1/2 + 1/2w)}{\Gamma(1/2)} -
\frac{2w}{1+w}\left(\frac{M}{E_{DE0}}\right)^{-2},
\end{eqnarray}
\end{widetext}
where we have used the fact that
$M/E_{DE0} >> 1$ to simplify the second term on the right-hand side.

We now use the constraint that observations require $w$ to be close to $-1$.
(Note that we do not take $w=-1$, as this would imply $M^4 = \rho_{DE0}$
and invalidate the entire model.  However, a value of $w$ even slightly less
than $-1$ allows for a phantom model with $M^4 \gg \rho_{DE0}$).
In the limit
where $w \rightarrow -1$, the numerator in Eq. (\ref{fraction2}) can be approximated
as
\be
\int_{r_1}^{r_2} \frac{r^{-\frac{2w+1}{2w}}}{\sqrt{1+r}}dr \approx 2 \ln \frac{\sqrt{r_2} +
\sqrt{1+r_2}}{\sqrt{r_1} + \sqrt{1+ r_1}}.
\ee
Further, we can simplify Eq. (\ref{denom}) in the limit where $w$ is close to
$-1$ (note that $\Gamma(z) \sim 1/z$ as $z \rightarrow 0$), to give
\be
\int_{0}^{r= (M/E_{DE0})^{4w/(1+w)}} \frac{r^{-\frac{2w+1}{2w}}}{\sqrt{1+r}} dr
\approx \frac{-2}{1+w}\left[1 - \left(\frac{M}{E_{DE0}}\right)^{-2}\right],
\ee
and our final expression for the coincidence fraction becomes
\be
\label{ffinal}
f \approx -(1+w) \ln \frac{\sqrt{r_2} +
\sqrt{1+r_2}}{\sqrt{r_1} + \sqrt{1+ r_1}}\biggl/[1-(M/E_{DE0})^{-2}].
\ee
The corresponding expression for the case of a phantom model with a future singularity
is identical to Eq. (\ref{ffinal}) without the $(M/E_{DE0})^{-2}$ in the denominator.
This difference is negligible as long as $M \gg E_{DE0}$, as it must be in
any reasonable cyclic phantom model.  This is just another
way of saying that the time needed for the universe to expand from the energy
scale $M$ to a future singularity is negligible compared to the time
for the expansion up to $M$.  Thus, the value for $f$ in the cyclic phantom
models is nearly identical to its value in models with a future singularity, and both
are given (for $w$ close to $-1$) by
\be
\label{fmain}
f \approx -(1+w) \ln \frac{\sqrt{r_2} +
\sqrt{1+r_2}}{\sqrt{r_1} + \sqrt{1+ r_1}}.
\ee
Equation (\ref{fmain}) is our main result.

As noted in Ref. \cite{Scherrer}, the exact values of $r_1$ and $r_2$ are not
well-defined, since the definition of a coincidence is somewhat arbitrary. 
However, if we require, for example, that the dark energy and dark matter densities be within
an order of magnitude of each other, then $r_1 = 1/10$ and $r_2 = 10$, yielding
$f = -1.56(1+w)$.  In this case, a coincidence fraction as large as $f = 0.1$
can be obtained for $w = -1.06$.  Thus, even for $w$ quite close to $-1$,
the oscillating phantom model provides a solution to the coincidence problem.

\section{Discussion}
The cyclic phantom model provides an attractive solution to the 
coincidence problem, since the universe spends an appreciable fraction,
$f$,
of each cycle in a state for which the dark energy and matter densities
are of the same order of magnitude.  For the models considered here,
we have shown that this fraction is essentially identical to the corresponding fraction in phantom
models with a big rip.  However, the cyclic phantom model provides a more
credible solution to the coincidence problem, in the sense that it does not entail a future singularity.
The cyclic phantom model has
the further advantage of unifying
inflation and dark energy.  (Indeed, that was the original motivation for this model).  Although
we have analyzed a generic toy model, these results apply, for example, to the model discussed in Ref. \cite{Ilie},
as long as this model is modified to include a matter component with the appropriate density.  In
Ref. \cite{Ilie}, the upper
and lower energy scales were taken to be $m \sim 1$ meV and $M \sim 10^{15}$ GeV, but as we have shown,
the value for $f$ is actually independent of $m$ and $M$ as long as $M \gg E_{DE0}$.

In the observationally allowed limit where $|1+w| \ll 1$, the coincidence
fraction $f$ is $-(1+w)$ times a constant of order unity.  Current
constraints on $w$ allow for a nonnegligible value for $f$.  However,
if future observations force $1+w$ to be sufficiently close to zero, this
scenario for resolving the coincidence problem (along with that outlined
in Ref. \cite{Scherrer}) will be ruled out.  Of course, these results
assume a constant value for $w$.  If one assumes a time-varying $w$,
then the value for $f$ can be larger than in constant $w$ models
\cite{Kujat}.

\section{Acknowledgments}
H.-Y. Chang was supported in part by the National Science
Foundation (DGE-0946822).
R.J.S. was supported in part by the Department of Energy (DE-FG05-85ER40226).

\end{document}